\documentstyle[preprint,aps,epsf]{revtex}
\tighten
\begin{document}
\newcommand{\rb}[1]{\raisebox{-1ex}[-1ex]{#1}}
\newcommand{\sst}[1]{\scriptscriptstyle{#1}}
\newcommand{\dfrac}[2]{\mbox{$\displaystyle\frac{#1}{#2}$}}

\draft

\title{
Pi-mesonic  decay of the hypertriton
}

\author{H.~Kamada\footnote{present address: Institut f\"ur
    Kernphysik, Fachbereich 5 der Technischen Hochschule Darmstadt, D-64289
    Darmstadt, Germany   },J.~Golak$^\dagger$, K.~Miyagawa$^\ddagger$,
         H.~Wita\l a$^\dagger$,
        W.~Gl\"ockle}
\address{Institut f\"ur Theoretische Physik II,
         Ruhr Universit\"at Bochum, D-44780 Bochum, Germany}
\address{$^\dagger$ Institut of Physics, Jagellonian University,
                    PL 30059 Cracow, Poland}
\address{$^\ddagger$ Department of Applied Physics,
                     Okayama University of Science,
                     Ridai-cho Okayama 700, Japan}

\date{\today}
\maketitle

\begin{abstract}
The $\pi$-mesonic decay 
of the hypertriton is calculated based on a hypertriton 
wavefunction and 3N scattering states, which are rigorous solutions of 3-body
Faddeev equations using realistic NN and hyperon-nucleon interactions. The
total 
$\pi$-mesonic decay rate is found to be 92\%  of the free 
$\Lambda$ decay rate, which is close to the experimental data. 
Together with the nonmesonic decay the total life time of $^{3}_{\Lambda}$H is
predicted to be 2.78 $\times$ 10$^{-10}$ sec which is
 6 \% larger than for the free $\Lambda$
particle.  
The differential decay rate is evaluated as a function of the pion momentum.
The decay into the $N+d+\pi$ channel is stronger than in the $3N+\pi$ channel
in contrast to the situation for the nonmesonic decay. The ratio for the decay
rate into $^{3}$He + $\pi^{-}$ to the decay rate 
into all channels including $\pi^{-}$
is found to be 0.40, which is close to the experimental value.
We visualise the decay into the dominant channel $p+d+\pi^{-}$ 
in a Dalitz plot. 
Finally we compare the polarisation of the outgoing proton in free unpolarised
$\Lambda$-decay to the polarisation of $^3$He in unpolarised $^3_\Lambda $H-decay 
and we compare the closely related
asymmetry of $\pi^-$ emitted parallel and antiparallel 
with respect to the spin-direction for a polarised $\Lambda$ to the corresponding
asymmetry for a polarised $^3_\Lambda$H.
\end{abstract}
\pacs{21.80.+a, 21.45.+v, 23.40.-s }

\narrowtext

\section{Introduction}
\label{secIN}

The hypertriton, a bound state of a proton, a neutron and a hyperon 
($\Lambda$ or $\Sigma$) is bound with respect to the 
$\Lambda -d $ threshold by  0.13$\pm$ 
0.05 MeV. 
We could reproduce that number\cite{1} by solving the Faddeev equations with
realistic NN forces and the Nijmegen hyperon-nucleon interaction \cite{2}. 
The hypertriton decays weakly into mesonic and nonmesonic channels.  The
nonmesonic decay channels are $^{3}_{\Lambda} $H$\to$ $d+n$ and
$^{3}_{\Lambda}$H$\to$$p + n + n $. We investigated them recently using
a realistic hypertriton wave function and realistic 3N continuum states and
evaluating the weak-strong transition $\Lambda N \to NN $ by exchange of
several mesons\cite{nonmesonic,Error}. 
We included $\pi,\eta,K,\rho,\omega$ and$ K^{*}$ meson exchanges.
They all contribute significantly, but there is a strong interference, which
leads to a result close to the one generated by the $\pi$-exchange only.
We found a total nonmesonic decay rate  of 6.39 $\times$ 10$^{7}$ sec$^{-1}$
\cite{nonmesonic,Error},
which is 1.7 \% of the free $\Lambda$-decay rate. These decays are neutron or
proton induced and we discussed that the corresponding total rates can not 
 be
separated experimentally. Under certain angular and energy restriction, they
can be separated, however.

In the mesonic decay mode there are more channels : $^{3}_{\Lambda} $H$\to \pi
^{-}(\pi^{0}) + {}^{3}He({}^{3}H)$,
 $^{3}_{\Lambda} $H$\to \pi
^{-}(\pi^{0}) + d + p(n)$ and $^{3}_{\Lambda}$H$\to \pi^{-}(\pi^{0}) + p + n +
p (n)$. In contrast to heavier hypernuclei, where mesonic decays are Pauli
blocked\cite{5}, here in the hypertriton they are by far the dominant
ones. Experimentally \cite{3} the life time 
of $^3_\Lambda$H ranges between (2.20 +1.02 -0.53) $\times$10$^{-10}$sec
  to (2.64 +0.92 -0.54) $\times$10$^{-10}$sec. 
  Further there are experimental data on the branching ratio $R=\Gamma 
  (^3_\Lambda H \to \pi^- + ^3He )$/$\Gamma ( ^3_\Lambda H \to $ all $\pi^-$ 
  meson modes ) ranging between 0.30 $\pm$ 0.07 \cite{3} to 0.39 $\pm$
  0.07\cite{Blok}. 

  The mesonic decay rates have been calculated using phenomenological 
  $^3_\Lambda$H and $^3$He wavefunctions by Dalitz\cite{Dalitz0}, Leon
  \cite{Leon} and by Kolesnikov and Kopylov \cite{Kole}, who used $^3_\Lambda$H
  and $^3$He wavefunctions found from variational calculations.
  More recently, again in a simple model the hypertriton and its mesonic decay 
  have been reconsidered by Congleton\cite{Congleton}.
   In view of the feasibility to perform rigorous 3-body calculations for 
   bound and continuum states based on modern baryon-baryon forces it appared 
   worthwhile to evaluate again  the mesonic decay channels using these 
   modern technical tools. Also we hope that these renewed considerations 
   based on modern forces will stimulate experiments on the $^3_\Lambda$H 
   decay modes.

In Section II we present our formalism. Our results are shown in section III. 
We summarise in section IV. Free $\Lambda$-decay properties and 
technical details are deferred to the Appendices.

\section{Formalism}
\label{secII}

The six mesonic decay channels of the hypertriton are not independent from
each other. According to the emprical $\Delta$I=${1\over 2}$ rule, which can
be realized by setting artificially the $\Lambda$ state to be $ \vert  t t_{z}
\rangle = \vert {1\over2} {-1\over 2} \rangle $ in isospin and introducing
$\vec \tau $ at  the vertex for $\Lambda \to N + \pi$ the following ratios for
decay rates result 
\cite{5}
\begin{eqnarray}
{ {\Gamma(^{3}_{\Lambda}H \to  \pi^{-} + ^{3}He) }
\over 
  {\Gamma(^{3}_{\Lambda}H \to  \pi^{0} + ^{3}H ) }  }= 
{ {\Gamma(^{3}_{\Lambda}H \to  \pi^{-} +  p  + d ) }
\over 
  {\Gamma(^{3}_{\Lambda}H \to  \pi^{0} +  n  + d ) } } = 
{ {\Gamma(^{3}_{\Lambda}H \to  \pi^{-} +  p + p + n )  }
\over 
  {\Gamma(^{3}_{\Lambda}H \to  \pi^{0} +  n + n + p ) } }
= 2
\label{ratio} 
\end{eqnarray}
Therefore we restrict ourselves to the $\pi^{-}$channels and introduce the
following notations: 
\begin{eqnarray}
\Gamma ^{He} &\equiv&{\Gamma(^{3}_{\Lambda}H \to  \pi^{-} + ^{3}He) }
\cr
\Gamma ^{p+d} &\equiv& {\Gamma(^{3}_{\Lambda}H \to  \pi^{-} +  p  + d ) }
\cr
\Gamma ^{p+p+n} &\equiv& {\Gamma(^{3}_{\Lambda}H \to  \pi^{-} +  p + p + n )
  }
{\rm .}
\end{eqnarray}
Then together with (\ref{ratio}) we get the full mesonic decay rate:
\begin{eqnarray}
\Gamma
 = {3 \over 2} \left( \Gamma ^{He} + \Gamma ^{p+d} +\Gamma ^{p+p+n}
\right)
\label{ratio2}
\end{eqnarray}

In the total momentum zero frame the three differential individual decay 
rates are:

\[
d \Gamma^{\rm He} \ = \ \frac12 \, \sum_{m \, m_{He} } \
\mid 
\sqrt{3}
< \Psi^{(-)}_{ {\vec k}_\pi \, {\vec k}_{He} \, m_{He} }
\mid {\hat O} \mid \Psi_{ {}^3_\Lambda H \, m } >  \mid^2 \,
\]
\begin{equation}
{{d {\vec k}_{\pi} } \over { 8 \pi^2 \omega_{\pi}} }
d {\vec k}_{He} \, 
\delta ( {\vec k}_\pi +  {\vec k}_{He} ) \,
\delta \left( M_{ {}^3_\Lambda H } - M_{^{3}He}\ -  \omega_{\pi} - \,
\frac { {\vec k}_{He}^{\ 2} }{6 M_N} \right) {\rm ,}
\label{e1}
\end{equation}
\[
d \Gamma^{\rm p+d} \ = \ \frac12 \, \sum_{m \, m_p \, m_d} \
\mid
\sqrt{3}
 < \Psi^{(-)}_{ {\vec k}_{\pi} \,
{\vec k}_p \, {\vec k}_d \,  m_p \, m_d }
\mid {\hat O} \mid \Psi_{ {}^3_\Lambda H \, m } >  \mid^2 \,
\]

\begin{equation}
{{d {\vec k}_{\pi} } \over { 8 \pi ^2\omega_{\pi}} }
d {\vec k}_p \, d  {\vec k}_d  \,
\delta ( {\vec k}_{\pi} + {\vec k}_p +  {\vec k}_d ) \,
\delta \left( M_{ {}^3_\Lambda H } - M_p - M_d - \omega_{\pi} - \,
\frac { {\vec k}_p^{\ 2} }{2 M_N} - \,
\frac { {\vec k}_d^{\ 2} }{4 M_N} \right)
\label{e2}
\end{equation}
and
\[
d \Gamma^{\rm p+p+n} \ = \ \frac12 \, \sum_{m \, m_1 \, m_2 \, m_3} \
\mid
\sqrt{3}
< \Psi^{(-)}_{ {\vec k}_{\pi} \, 
{\vec k}_1 \, {\vec k}_2 \, {\vec k}_3 \, m_1 \, m_2 \, m_3 }
\mid {\hat O} \mid \Psi_{ {}^3_\Lambda H \, m } >  \mid^2 \,
\]
\begin{equation}
{{d {\vec k}_{\pi} } \over { 8 \pi ^2 \omega_{\pi}} } 
d {\vec k}_1 \, d {\vec k}_2 \, d {\vec k}_3   \,
\delta ({\vec k}_{\pi}+ {\vec k}_1 + {\vec k}_2 + {\vec k}_3 ) \,
\delta \left( M_{ {}^3_\Lambda H } - 3 M_N \, - \omega_{\pi} \, -
\frac { {\vec k}_1^{\ 2} }{2 M_N} - \,
\frac { {\vec k}_2^{\ 2} }{2 M_N} - \,
\frac { {\vec k}_3^{\ 2} }{2 M_N} \right){\rm .}
\label{e3}
\end{equation}
Here $\Psi^{(-)}$ are appropriate pion- three-nucleon scattering states,
$ {\hat O} $ the vertex operator ($\Lambda \to \pi^{-} + p $),  
$\Psi_{ {}^3_\Lambda H}$
the hypertriton wavefunction and $ \omega_{\pi} = \sqrt{
 m_{\pi}^{2}  
+ {\vec k}_{\pi}^{2} } $. 
The scattering states $\Psi^{(-)}$  are normalized to $\delta$-functions with
respect to the asymptotic momenta of relative motions. 
The factor $\sqrt{3}$ results from antisymmtrisation \cite{Dalitz} with
respect to the nucleons, which are treated as identical in the isopin
formalism. 
Except for the choice of the relativistic energy $\omega_{\pi}$ of the pion
our calculation is nonrelativistic.
The binding energies $\epsilon$ are defined as usual in terms of the nucleon
($M_{N}$) and $\Lambda$ mass ($M_{\Lambda}$) :

\begin{eqnarray}
  M_{ {}^3 He} \, &=& \, 3 M_N  + \epsilon_{^{3}He}  \cr 
  M_{ {}^3_\Lambda H} \, &=& \, 2 M_N + M_\Lambda + \epsilon_{^{3}_{\Lambda}H}  \cr
  M_d \, &=& \, 2 M_N + \epsilon_d 
\label{e4}
\end{eqnarray}
The individual momenta in (\ref{e3}) are connected to 
Jacobi momenta as
\begin{eqnarray}
 & {\vec p} \, = \, {1 \over 2} \ ({\vec k}_1 - {\vec k}_2) \nonumber \\[4pt]
 & {\vec q} \, = \, {2 \over 3} \ ({\vec k}_3 -
{1 \over 2} \ ({\vec k}_1 + {\vec k}_2))
\label{e5}
\end{eqnarray}
The momenta in (\ref{e2}) are a special case with $\vec k  _{3 } =\vec k_{p}$
and $\vec k_{1} + \vec k_{2} = \vec k_{d}$.
In (\ref{e1}) the Jacobi momenta for $^{3}$He are like in Eq.(\ref{e5}), but
of course referring to the internal motion of $^{3}$He.
For the hypertriton we use a representation based on the Jacobi momenta 
\begin{eqnarray}
 & {\vec p} \ ' \, = \, {1 \over 2} \ ({\vec k}_1 \ '  - {\vec k}_2 \ ') \nonumber \\[4pt]
 & {\vec q} \ ' \, = \, 
{ {2 M_{N} {\vec k}_{3} \ ' - M_{\Lambda} ({\vec k}_{1} \ ' +{\vec k}_{2} \ ')} \over
{ 2 M_{N} + M_{\Lambda}} }
\label{jacobi2}
\end{eqnarray}
Eqs. (\ref{e1}-\ref{e3}) refer to the total momentum zero frame and thus 
the individual momenta for the hypertriton obey of course 
${\vec k}_{1} \ ' + {\vec k}_{2} \ ' + {\vec k}_{3} \ ' 
=0 $. Further 
we selected particle 3 to be the $\Lambda$ particle.

The phase space factors are easily evaluated. 
The one for $\pi^{-}$-$^{3}$He decay is:   
\begin{eqnarray}
\rho^{He}
 &\equiv& {1 \over { 8 \pi^{2}\omega _{\pi}}} \int d \vec k _{\pi} d \vec k
_{^{3}He} 
\delta ( {\vec k}_\pi +  {\vec k}_{He} ) \,
\delta \left( M_{ {}^3_\Lambda H } - M_{^{3}He}\ -  \omega_{\pi} - \,
\frac { {\vec k}_{He}^{\ 2} }{6 M_N} \right)
\cr
&=&
    {1 \over {8 \pi ^{2} } } {{3M_{N} k_{\pi} } \over {3M_{N} + \omega_{\pi}}}
d \hat {k_{\pi}}
\label{rhohe}
\end{eqnarray}
where $\vert \vec k_{\pi} \vert$ is kinematically fixed as 
\begin{eqnarray}
k_{\pi} = \vert \vec k _{\pi} \vert =
{\sqrt{(M_{^{3}_{\Lambda}H}^{2}+M_{^{3}He}^{2}-m_{\pi}^{2})^{2} - 4
  M_{^{3}_{\Lambda}H}^2 M_{^{3}He}^2 } \over {2 M_{^{3}_{\Lambda}H}} } 
\end{eqnarray}
The numerical value of $\vert \vec k_{\pi} \vert $ given by 
this exact relativistic expression agrees within 0.022 \%
with the corresponding number obtained from a 
corresponding nonrelativistical formula.
The ones for $p + d + \pi ^{-}$ and $p + p + n + \pi ^{-}$ decays for fixed
$\vec k _{\pi}$ are similar to the ones given in \cite{nonmesonic}.

The exact treatment of the final continuum states in Eqs.(\ref{e1}-\ref{e3})
requires the solution of a 4-body problem, which is beyond our present
capability. Therefore we neglect the final state interaction of the pion.
For its treatment in the framework of an optical potential we refer to a
recent article \cite{Kelkar}. Under the approximation of a free pion the
matrix elements in Eqs.(\ref{e1}-\ref{e3}) shrink to ones  which refer only to
three baryons. The resulting operator $O$ acting in that reduced space,
related to the elementary  process $\Lambda \to p + \pi^{-}$ is given in a
relativistic notation as \cite{5}:  
\begin{eqnarray}
 O = i\sqrt{2} G_{F} m_{\pi}^{2} F(\vec k_{\pi})
 \bar u_{N} (\vec k_{3})  
(A_{\pi} + B_{\pi} \gamma_{5}) u_{\Lambda} ( \vec k_{3}) 
\label{eo0}
\end{eqnarray}
where $\bar u$,  $u$ are  Dirac spinors and  $G_{F} m_{\pi}^{2}$=2.21$\times$
10$^{-7}$  the weak coupling constant. The constants $A_{\pi}$=1.05 and
$B_{\pi}$ = -7.15 measure the parity violating and conserving
parts\cite{4,nonmesonic}.   
The factor $\sqrt{2}$ arises from the "spurion" character of $\Lambda$ and
from $\vec \tau \cdot \vec \phi$, where $\vec \phi$ is the isovector pion
field. In case of the $n+\pi^{0}$ decay that factor is (-1).
In nonrelativistic reduction the simple operator results
\begin{eqnarray}
 O \longrightarrow i \sqrt{2} G_{F} m_{\pi}^{2} F(\vec k_{\pi}) 
(A_{\pi} + {{ B_{\pi} } \over { 2 \overline{ M}} }
 \vec \sigma  \cdot
\vec k_{\pi} ) 
\label{eo}
\end{eqnarray}
with $\overline{M} \approx (M_{N}+M_{\Lambda})/2$.
The form factor  $F(\vec k_{\pi})$ is chosen of the monopole type and 
is given in
Appendix A together with a brief display of the free
$\Lambda$-decay rate. 
The use of the nonrelativistic reduction (\ref{eo}) introduces only
a 2\% shift for that decay rate. 
Finally using Eqs. (\ref{e5}),(\ref{jacobi2}), the $\Lambda$ rest frame  condition
and our choice for the $\Lambda$ to be  particle 3 one has 
\begin{eqnarray}
\vec k _{\pi} = {3 \over 2} ( \vec q \ ' - \vec q )
\end{eqnarray}

\begin{figure}
\centering
\mbox{\epsfysize=50mm \epsffile{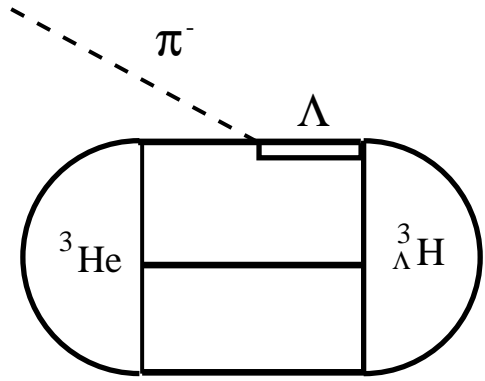}} \vspace*{4mm}
\caption{The nuclear matrix element  for the process $^3_\Lambda$H
 $ \to \pi ^{-} + {}^3$He}
\label{f1}
\end{figure}

\begin{figure}
\centering
\mbox{\epsfysize=50mm \epsffile{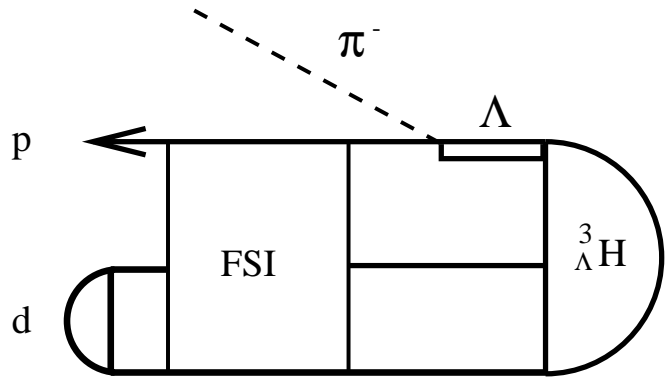}} \vspace*{4mm}
\caption{The nuclear matrix element  for the process 
 $^3_\Lambda$H
$\to \pi^- + p + d $}
\label{f2}
\end{figure}

\begin{figure}
\centering
\mbox{\epsfysize=50mm \epsffile{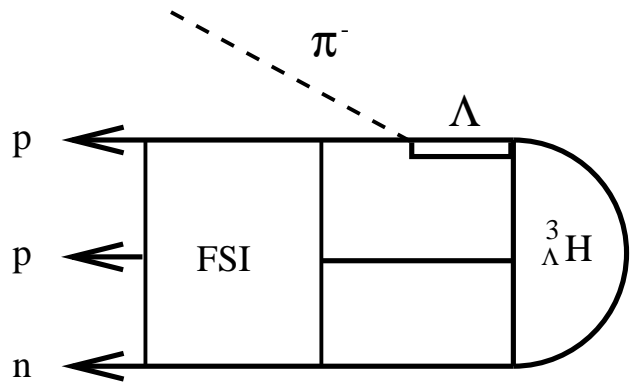}} \vspace*{4mm}
\caption{The nuclear matrix element  for the process  $^3_\Lambda$H $\to
\pi^- + p + p + n $ }
\label{f3}
\end{figure}

Though we neglect the interaction of the pion with the 3 nucleons 
we treat the
one for the 3 final nucleons exactly.
This is depicted graphically in Figs.\ref{f1}-\ref{f3}. 
The final state interaction  among the 3 nucleons 
can be performed in analogy to electron scattering on ${}^3$He~\cite{9}.
We exemplify it for the nnp breakup process. For our notation in general
we refer to~\cite{10}.

The 3N scattering state expressed in Jacobi momenta, $\Psi^{(-)} \equiv
\Psi^{(-)}_{\vec p \vec q m_{1}m_{2}m_{3}}$, is Faddeev decomposed 
\begin{equation}
\Psi^{(-)} \, = \, (1 + P ) \psi^{(-)}{\rm ,}
\label{e9}
\end{equation}
where $P$ is the sum of a cyclical and anticyclical permutation of 3 objects
and $ \psi^{(-)}$ is one Faddeev component.
It obeys the Faddeev equation
\begin{equation}
\psi^{(-)} \, = \, \phi^{(-)} + G_0^{(-)} t^{(-)} P  \psi^{(-)}
\label{e10}
\end{equation}
with
\begin{equation}
\phi^{(-)} \, = \, ( 1 + G_0^{(-)} t^{(-)} ) \phi^{a}_0
\label{e11}
\end{equation}
and
\begin{equation}
\phi^a_0 \, = \, \frac1{\sqrt{3 !}} ( 1 - P_{12} ) \mid \phi_0 > \, \equiv \,
\frac1{\sqrt{6}} ( 1 - P_{12} ) \mid {\vec p} > \mid {\vec q} >
\label{e12}
\end{equation}
Here $ G_0^{(-)}$ is the free three-nucleon  propagator,
$ t^{(-)}$ the NN (off-shell) t-matrix and $\frac1{\sqrt{6}}$
takes care of the identity of the three nucleons. Note that
$P_{12}$ acts in the two-body subsystem described by the relative momentum
$\vec p$.
As shown in  \cite{nonmesonic} the nuclear matrix element for the 3N breakup
can be written as :
\begin{equation}
<  \Psi^{(-)}_{ {\vec p} \, {\vec q} \, m_{1} \, m_{2} \, m_{3}} \mid
 { O}  \mid \Psi_{ {}^3_\Lambda H } > \, = \,
 < \phi^a_0 \mid (1 + P) { O}  \mid \Psi_{ {}^3_\Lambda H } > \, + \,
 < \phi^a_0 \mid (1 + P) \mid U > {\rm ,}
\label{e14}
\end{equation}
where $ \mid U >$ obeys the Faddeev equation
\begin{equation}
 \mid U > \, = \, t G_0 (1 + P) { O}  \mid \Psi_{ {}^3_\Lambda H } >
\, + \,  t G_0 P \mid U >
\label{e15}
\end{equation}
The action of the operator $O$ will be described 
in the Appendix B.
The hypertriton state has a $\Lambda NN$ and a $\Sigma NN$  part. Though the
$\Lambda -\Sigma$ conversion is crucial for the binding of the hypertriton, the
$\Sigma NN$ admixture is very small \cite{1} and we neglect it. Thus
we also neglect the contribution of the $\Sigma$-decay, as we also did in the
case of the nonmesonic decay\cite{nonmesonic}.

In ~\cite{1} the hypertriton state has been determined in a partial wave
representation and we refer to~\cite{1} for the details of our notation.
Here we need only the form
\begin{eqnarray}
\mid \Psi_{ {}^3_\Lambda H } >  \, = \,
\sum_{\alpha} \, \int d \, p \, p^2 \,
\int d\, q \, q^2 \, \mid p q \alpha > \, \Psi_{\alpha} ( p q ) {\rm ,}
\end{eqnarray}
where $p, q $ are the magnitudes of the Jacobi momenta (\ref{jacobi2})
 and $\alpha$
denotes the following set of discrete quantum numbers
\begin{eqnarray}
\alpha \, \equiv \,
( l s ) j \, (\lambda \frac12 ) I ( j I ) J (t 0 ) T
\label{qnum}
\end{eqnarray}
Here $( l s ) j$ describes the coupling of orbital angular momentum $l$
and total spin $s$ to the total two-body angular momentum $j$ of
the NN subsystem, $ (\lambda \frac12 ) I $ the corresponding
coupling of orbital and spin angular momentum  of $\Lambda$ to its
total angular momentum $I$, $  ( j I ) J $, the resulting $jI$ coupling
to the total angular momentum $J$ and finally the isospin coupling of
the two-nucleon isospin $t= 0 $ and the isospin zero of the $\Lambda$ particle 
to total isospin $T= 0$.

Also for the evaluation of the matrix elements in (\ref{e14})
and the solution
of the Faddeev equation~(\ref{e15}) we work in a partial wave representation,
using a complete set of basis states now for three nucleons.
They are again denoted as $  \mid p q \alpha >_N $ but
adding a subscript $N$ to indicate that the Jacobi momenta are now
from (\ref{e5}). Furthermore one has to note that this is a subset
of states antisymmetrized in the subsystem of particles 1 and~2,
thus $ ( l + s + t ) $ has to be odd.

Now projecting the Faddeev equation into the basis  $  \mid p q \alpha >_N $
and inserting appropriate decompositions of the unity one gets
\[
{}_N < p q \alpha \mid U > \, = \,
\]
\[
\sum \hspace{-0.5cm} \int \,
\sum \hspace{-0.5cm} \int  \,
{}_N < p q \alpha \mid  t G_0 (1 + P ) \mid p' q' {\alpha}' >_N \,
{}_N < p' q' {\alpha}' \mid  { O} \mid p'' q'' {\alpha}'' > \,
\Psi_{{\alpha}''} ( p'' q'') \ + \
\]
\begin{equation}
\sum \hspace{-0.5cm} \int  \,
{}_N < p q \alpha \mid  t G_0 P \mid p' q' {\alpha}' >_N \,
{}_N < p' q' {\alpha}' \mid  U >
\label{e29}
\end{equation}
This is a coupled set of integral equations, with a kernel part,
which is well known~\cite{8} from 3N scattering, and an inhomogeneous
term, whose part left of $ { O} $ is also familiar from
electron scattering~\cite{9}. What is left as a new structure
is the application of the
$ { O} $-matrix onto the wavefunction component of the
hypertriton. 
This is given in Appendix B.

Once the amplitudes
$ {}_N \! < p q \alpha \mid U > $ are determined,
the matrix element in (\ref{e14}) 
 is  evaluated by quadrature in the manner described in~\cite{8}
and references therein.
The first matrix element in Eq. (\ref{e14}), 
the plane wave impulse approximation with
respect to the nucleons, is also evaluated by the same techniques via partial
wave decomposition.

\section{Results}
\label{secIII}

We used a hypertriton wave function based on the Nijmegen~93 NN
potential~\cite{12}
and the Nijmegen YN interaction~\cite{2}, which include
the $\Lambda - \Sigma$
transitions. The number of different $\alpha$ quantum numbers (\ref{qnum}) 
, usually called channels,  used in the
solution of the corresponding Faddeev equation is 102. This leads to a fully
converged state, which has the proper antisymmetrisation among the
two nucleons built in. Also the NN and YN correlations are
exactly included  as generated by the various baryon-baryon forces
(see~\cite{1}). The $\Sigma NN$ part of the state has a probability
of 0.5 \% and will be neglected.

Let us first regard the decay channel: 
\[
{}^{3}_{\Lambda}H \to \pi ^{-} + {}^{3}He
\] 
In that  channel we use the ${}^{3}He$
wave function generated by the 
Faddeev equation with 
 the Nijmegen 93  NN
interaction~\cite{12}. The kinematically fixed value 
of the pion momentum is  
$k_{\pi}$=117.4MeV/c.
From  (\ref{e1}) and (\ref{rhohe}) we obtain the total decay rate 
$\Gamma^{He}$=9.36$\times$ 10$^{8}$ sec$^{-1}$ as our
theoretical prediction. It should be compared to the value  8.84$\pm$2.27
$\times$10$^{8}$ sec$^{-1}$ which we estimate from the total  $^{3}_\Lambda$H
    decay life time $\tau=$( 2.64 +0.92-0.54) $\times$10$^{-10}$
  \cite{3} ( neglecting the nonmesonic piece), using
   the factor 3/2 of Eq.(\ref{ratio2}) 
from isospin and the ratio R (see Introduction). 
The decay rates and life times for both  mesons are given in Table \ref{t1}.

Let us now ask more detailed questions in relation to that $^3$He channel. 
For the free $\Lambda$-decay into $ p + \pi^-$ the proton turns out to be 
polarised as a consequence of the interference between the two operators in 
(\ref{eo}). 
 For unpolarised $\Lambda$'s the polarisation of the proton in the 
direction of the outgoing $\pi^-$ is easily evaluated as 
\begin{eqnarray}
P_{\Lambda} &=& { { \sum _{m_\Lambda} \sum_{m_p}  m_p \vert 
( A_\pi + { B_\pi \over {2 \overline M}} \vec \sigma \cdot \vec k _\pi )
_{m_p m_\Lambda} \vert ^2   } 
\over 
 { \sum _{m_\Lambda} \sum_{m_p} \vert 
( A_\pi + { B_\pi \over {2 \overline M}} \vec \sigma \cdot \vec k _\pi )
_{m_p m_\Lambda} \vert ^2   }  }
\cr
&=&
{ { { { A_{\pi} B_{\pi} } \over { 2 \overline M}} \vert \vec k_{\pi} \vert  }
\over
{ A_{\pi} ^{2}   +  \left( { {B_{\pi}}  \over { 2 \overline M } } \right) ^{2} \vec 
k_{\pi}^{2}   }
}
\label{pol1}
\end{eqnarray}
$P_{\Lambda}$ is related to the measured quantity $\alpha$ = 
$-$0.64$\pm$ 0.01
\cite{kelley} (factor 2) 
 and agrees of
course with (\ref{pol1}), using our values for $A_{\pi}$ and $B_{\pi}$.
It results $P_\Lambda$ = -0.322.

Now back to the $^{3}$He-channel we ask for the polarization of $^{3}$He 
in the direction of $\pi^{-}$ for an unpolarised $^{3}_{\Lambda}$H.
It is given as 
\begin{eqnarray}
P_{^{3}{He}} &=&
{ { \sum _{m} \sum_{m'}  m'  \vert \sqrt{3} 
\langle \Psi ^{(-)} _{\vec k_{\pi} \vec k_{^{3}{He}}  } m ' \vert \hat 
O \vert \Psi _{^{3}_{\Lambda}{H} } m \rangle \vert ^{2 } 
}
\over 
 { \sum _{m} \sum_{m'}   \vert \sqrt{3} 
\langle \Psi ^{(-)} _{\vec k_{\pi} \vec k_{^{3}{He}}  } m ' \vert \hat 
O \vert \Psi _{^{3}_{\Lambda}{H} } m \rangle \vert ^{2 } 
}
}
\end{eqnarray}
We find $P_{^{3}\mbox{He}}$ = 0.134, which has the opposite sign of $P_\Lambda$.
This finds a simple explanation in regarding a further observable.
This is the difference in the probability for 
 $\pi^{-}$s leaving in
the direction of a polarised $^{3}_{\Lambda}$H to the $\pi^{-}$ leaving
opposite to that direction. This quantity can be compared to the one for the
free $\Lambda$-decay of a polarised $\Lambda$. 
In a polarised hypertriton the $\Lambda$ has a small polarisation
$p_{\Lambda}$= $-$0.166 \cite{1}. 
Therefore we expect a change of sign between the 
two differences and a change of
the magnitude because of the nuclear wavefunctions.
A simple calculation for the free $\Lambda$-decay is 
\begin{eqnarray}
A^{\Lambda} &=&
{ { {d \Gamma ^{\Lambda}} \over d \hat k _{\pi }  } \vert _{\theta_{\pi} = 0 } - 
 { { d \Gamma ^{\Lambda}} \over d \hat k _{\pi }  } \vert _{\theta_{\pi} = \pi } 
    } \over 
{ { {d \Gamma ^{\Lambda}} \over d \hat k _{\pi }  } \vert _{\theta_{\pi} = 0 } + 
 { { d \Gamma ^{\Lambda}} \over d \hat k _{\pi }  } \vert _{\theta_{\pi} = \pi } 
    }
\cr
&=& 2 P_{\Lambda}
\end{eqnarray}
Here $\theta_{\pi}$ is the angle of the emitted $\pi^{-}$ in relation to the 
direction of the $\Lambda$ polarisation.

The corresponding quantity of the $^{3}_{\Lambda}$H-decay  into $\pi^{-}$ 
+ $^{3}$He is 
\begin{eqnarray}
A^{^{3}_{\Lambda}{H}} &=& 
{ { {d \Gamma ^{He}} \over d \hat k _{\pi }  } \vert _{\theta_{\pi} = 0 } - 
 { { d \Gamma ^{He}} \over d \hat k _{\pi }  } \vert _{\theta_{\pi} = \pi } 
    } \over 
{ { {d \Gamma ^{He}} \over d \hat k _{\pi }  } \vert _{\theta_{\pi} = 0 } + 
 { { d \Gamma ^{He}} \over d \hat k _{\pi }  } \vert _{\theta_{\pi} = \pi } 
    }
\cr
&=&
{ {   \sum _{m'}  \vert \langle \Psi ^{(-)}  _{\vec k _{\pi }  \vec k
      _{^{3}He} } m' \vert \hat O \vert \Psi_{^{3}_{\Lambda}H } m={1\over2}
    \rangle \vert ^{2 } _{\theta_{\pi} = 0 }   - 
 \sum _{m'}  \vert \langle \Psi ^{(-)}  _{\vec k _{\pi }  \vec k
      _{^{3}He} } m' \vert \hat O \vert \Psi_{^{3}_{\Lambda}H } m={1\over2}
    \rangle \vert ^{2 } _{\theta_{\pi} = \pi } 
} \over 
 {   \sum _{m'}  \vert \langle \Psi ^{(-)}  _{\vec k _{\pi }  \vec k
      _{^{3}He} } m' \vert \hat O \vert \Psi_{^{3}_{\Lambda}H } m={1\over2}
    \rangle \vert ^{2 } _{\theta_{\pi} = 0 }   + 
 \sum _{m'}  \vert \langle \Psi ^{(-)}  _{\vec k _{\pi }  \vec k
      _{^{3}He} } m' \vert \hat O \vert \Psi_{^{3}_{\Lambda}H } m={1\over2}
    \rangle \vert ^{2 } _{\theta_{\pi} = \pi } }
}
\end{eqnarray} 
A detailed look into the expressions (\ref{A3}) and (\ref{A4}) 
given in Appendix B and the
corresponding ones where  $\hat k_{\pi}$ is opposite to the z-direction,
reveals that 
\begin{eqnarray}
A^{^{3}_{\Lambda}{H}} = 2 P_{^{3}{He} }
\end{eqnarray}
Consequently also $P_{^3He}$ should be opposite in sign to $P_\Lambda$.

Let us now investigate the breakup channels:
\[
{}^{3}_{\Lambda}H \to \pi ^{-} + p + d ~~~~~\mbox{and}~~~~~~
{}^{3}_{\Lambda}H \to \pi ^{-} + p + p + n  \]
For these channels we must  solve the Faddeev equation (\ref{e29}).
Again we use the realistic Nijmegen 93 NN interaction\cite{12}.
The technical steps are the same 
as in the  case of the nonmesonic decay \cite{nonmesonic}.
In table \ref{t1} we show 
our theoretical predictions for the summed up rates ($\pi^-$ and $\pi^0$)  
into the deuteron and 3N channels which 
together with  the rates into the 3N bound states 
given above  leads to the 
 total rate 3.5 $\times$ 10 $^{9}$ sec$^{-1}$.
It results a 
theoretical life time with respect to mesonic decay only of  $\tau$ =  2.9 
 $\times$10$^{-10}$ sec. 
We see that the strongest decay goes into the $p+d$ channel followed by the
transition into the 3 nucleon bound state $^{3}$He. Both are  much
stronger than the decay into the 3N channel. For the nonmesonic decay
also shown in Table \ref{t1} 
this is different. There the $n+n+p$ channel is the dominant one.
Our results from\cite{nonmesonic}
are reproduced for the convenience of the reader. Please note that they are
larger by a factor of 3 due to an overlooked factor $\sqrt{3}$ resulting from
correct antisymmetrisation\cite{Error}. 
The total theoretical lifetime with respect to all decay channels turns 
out to be 2.78 $\times$10$^{-10}$ sec which is not far from the 
largest value 
 ( 2.64 + 0.92 - 0.54 ) $\times$ 10$^{-10}$ sec in the range of experimental 
  data \cite{3}.
The theoretical ratio  $R = {{\Gamma^{He}} /( {\Gamma ^{He} + \Gamma^{p+d}
    +\Gamma^{p+p+n} }) }$  is  0.40
which is in reasonable  agreement with the experimental mean value of 
0.35 $\pm$ 0.04 \cite{Congleton}. 

\begin{figure}
\centering
\mbox{\epsfysize=100mm \epsffile{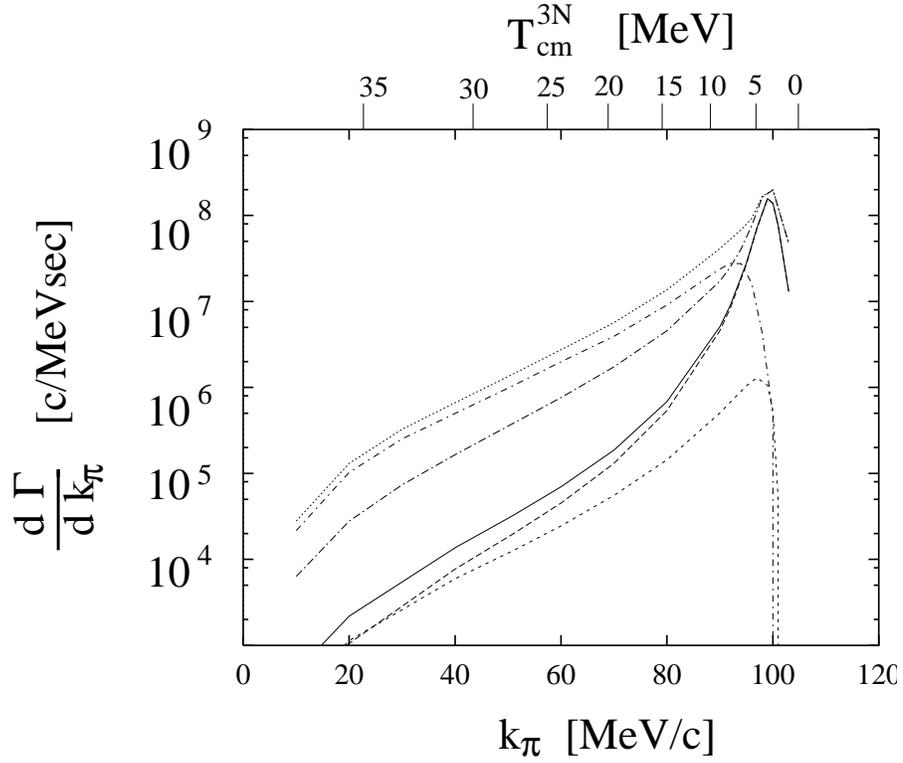}} \vspace*{4mm}
\caption{
Differential decay rates ${{d \Gamma ^{p+d} } \over {d k_{\pi}}}$ (long dashed
curve ), ${{d \Gamma ^{p+p+n} } \over {d k_{\pi}}}$ (short dashed curve) and
their sum (solid curve) including FSI. Neglecting FSI the rates are
drastically shifted :${{d \Gamma ^{p+d} } \over {d k_{\pi}}}$ (long dashed
dotted ) , ${{d \Gamma ^{p+p+n} } \over {d k_{\pi}}}$ (short dashed dotted)
and their sum (dotted). } 
\label{f5}
\end{figure}

\begin{figure}
\centering
\input{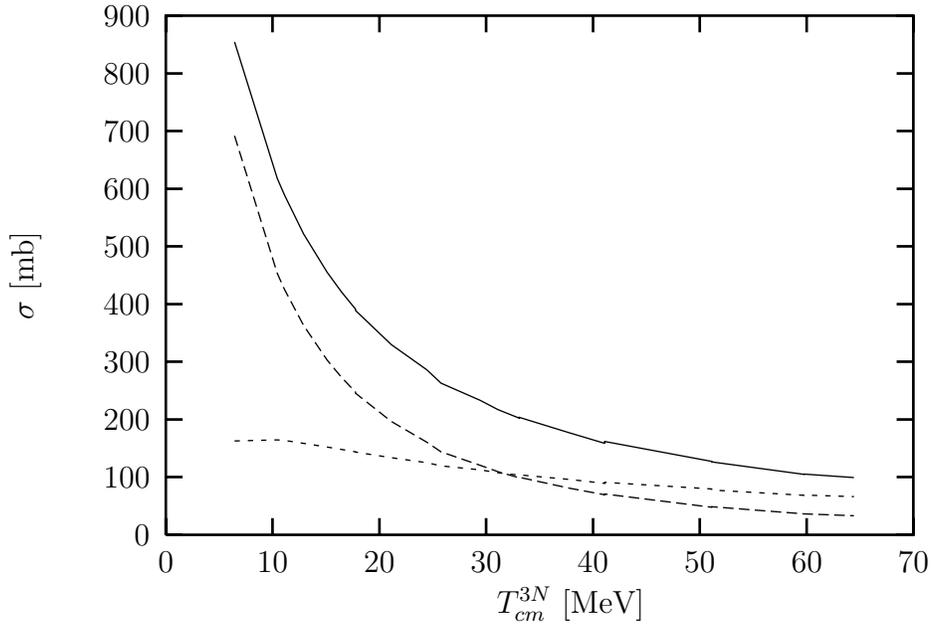}
\bigskip
\bigskip
\caption{ Angular integrated  cross sections for 3N scattering: Total nd coss
  section (solid curve), total elastic cross section (dashed curve), total
  breakup cross section (dotted curve).}
\label{f5.2}
\end{figure}

As additional information we show in Fig.\ref{f5} the differential decay rates
$d \Gamma ^{p+d}$ / $d k_{\pi}$ and $d \Gamma ^{p+p+n}$ / $d k_{\pi}$ as well
as their sum.
These quantities result by integrating over all variables except for
$k_{\pi}$.
 Both individual rates peak near the $\pi^{-}+ p +p +n$
threshold at $k _{\pi}$ = 101.3 MeV/c. 
The rate into the p+d channel dominates. 
It is only at $k_{\pi}$ about 20MeV/c, that the 3N channel is equally strongly
populated and overtakes the p + d channel for even smaller pion momenta.
In Fig. \ref{f5} we also display the 3N c.m. energy $T_{cm}^{3N}$,
which is kinematically
connected to the pion momentum. At $k_{\pi} \approx$ 20MeV/c it reaches
$T_{cm}^{3N}$=35MeV. 
It is around this energy where the total breakup cross section in n + d
scattering also overtakes the total elastic n + d cross section. This is shown
in Fig. \ref{f5.2}. It is therefore tempting to interprete the outcome in
Fig. \ref{f5} to result from the scattering of the nucleon arising from the
weak $\Lambda$ decay from the deuteron in the hypertriton. At low
c.m. energies $T_{cm}^{3N}$ elastic scattering of that nucleon from the
deuteron dominates and around $T_{cm}^{3N}$=35MeV 
the breakup process catches up. 
The stronger energy dependence in Fig. \ref{f5} in comparison to Fig. \ref{f5.2}
is caused by the production process of the nucleon out of the $\Lambda$-decay.
If one switches off the final state interaction between the proton and
the deuteron the decay rates are drastically shifted. Also then the three
nucleon breakup dominates except near the highest pion energy. 
 
It is conceivable that Coulomb force effects in the $\pi^{-}$ channel, where a
proton scatters off the deuteron will influence the rates. 
 The elastic scattering in the  p d channel 
 is stronger than in the nd
channel, which we calculated.
We neglected the p p Coulomb force totally.
 This is of course a quantitative question, which
should be checked in the future in a fullfledged 3N continuum calculation
including the Coulomb force.

\begin{figure}
\centering 
\mbox{\epsfysize=120mm \epsffile{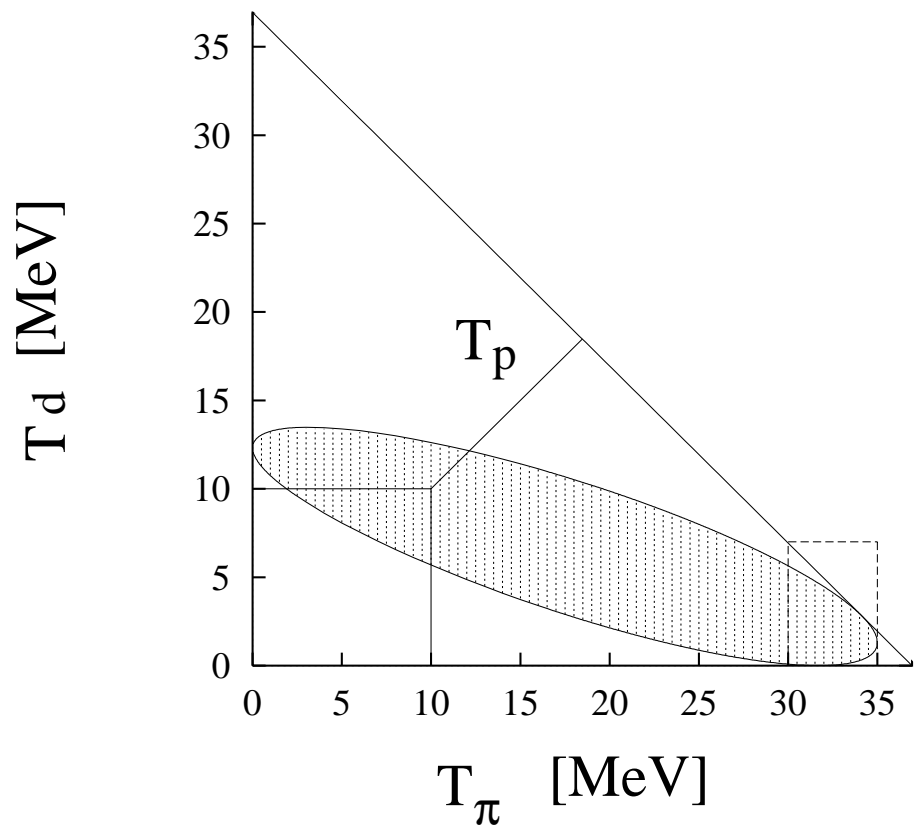}} \vspace*{4mm}
\bigskip
\caption{A triangle chosen for the Dalitz plots in Figs.\ref{f7} and \ref{f10}. The
  kinematically allowed events lie in the shaded area. Nearly all events
  occur at the right end, in the subdomain encircled by a dashed line.}
\label{f6}
\end{figure}

Finally we show the energy distribution of the 
meson, the nucleon and the deuteron in form of a Dalitz plot.
The triangle chosen for the Dalitz plot is shown 
in Fig. \ref{f6}.
The quantity to be presented is $ { {d \Gamma } \over {d T_{\pi} d T_{d}} } $
which results from Eq. (\ref{e2})
by integrating over all angles. 
We get 
\begin{eqnarray}
{ {d \Gamma } \over {d T_{\pi} d T_{d}} } 
= { 1\over 2} \sum _{m m_{p} m_{d}}  \int d\phi_{d} d \hat k_{\pi} 
\vert \sqrt{3} \langle \Psi ^{(-)} _{\vec k_{\pi} \vec k_{p} \vec k_{d} }
\vert  \hat O \vert \Psi _{^{3}_{\Lambda}H} m \rangle \vert ^{2 }
{ { M_{N} ^{2} } \over {4 \pi ^2}} 
\label{dali}
\end{eqnarray}
where cos $\theta_{d}$ is kinematically fixed. 
After summation over the spin magnetic quantum numbers of $^{3}_{\Lambda}$H,
the proton and the deuteron and using the momentum conserving
$\delta$-function, the matrix element squared depends only on the angle
$\theta_{d}$ between $\hat k_{\pi}$ and $\hat k_{d}$. Therefore the angular
integrations in (\ref{dali}) are trivial and lead just to a factor $8\pi^{2}$.

\begin{figure}
\centering
\input{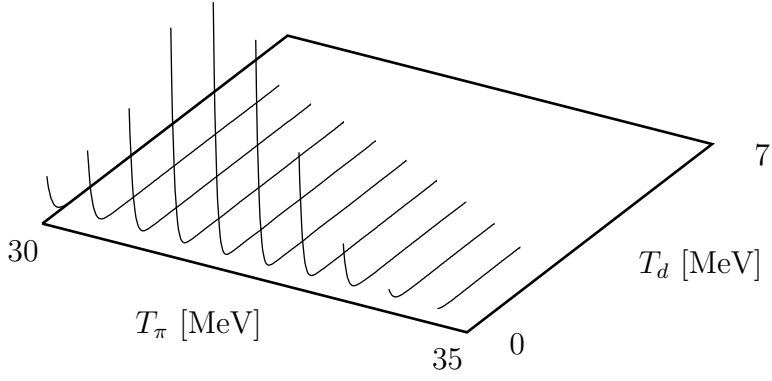}
\bigskip
\caption{The distribution ${{ d \Gamma ^{p+d}} \over {d T_{\pi} d T_{d}}}$ 
in the subdomain of Fig. \ref{f6}. }
\label{f7}
\end{figure}

\begin{figure}
\centering
\mbox{\epsfysize=100mm \epsffile{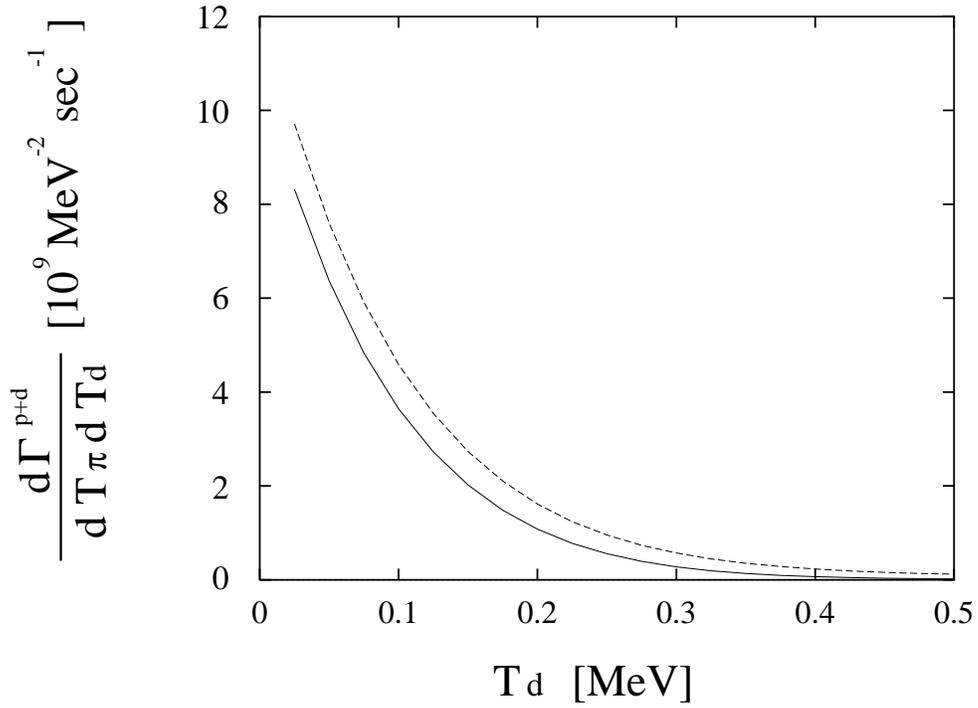}} \vspace*{4mm}
\bigskip
\bigskip
\caption{The distributions ${{ d \Gamma ^{p+d}} \over {d T_{\pi} d T_{d}}}$ 
including FSI (solid curve) and neglecting FSI (dashed curve) for
 fixed $T_{\pi }$ = 32MeV. 
 Note that the curves for that $T_{\pi}$ do not
 start at $T_{d}$ = 0. }
\label{f8}
\end{figure}

The three kinetic energies 
\begin{eqnarray}
&{}& T_{\pi } = \sqrt{ m_{\pi } ^{2} + {\vec k_{\pi}}^{2}   } - m_{\pi} \cr
&{}& T_{p } = {{ \vec k_{p}}^{2} \over { 2 M_{N}} } \cr 
&{}& T_{d } = {{\vec k_{d}}^{2} \over { 4 M_{N}} }
\end{eqnarray}
with $\vec k_{\pi} + \vec k_{p } + \vec k_{d} = 0 $ sum up to the total
kinetic energy $T_{cm}$= 36.9 MeV. 
As is well known the kinetic energies can be read off as the
perpendicular distances to the sides of the triangle as depicted in
Fig. \ref{f6}. 
The kinematically accessible events have to lie in the shaded area shown in
Fig. \ref{f6}.   It turns out that essentially all events concentrate in the
corner encircled by the dashed line in Fig. \ref{f6}. The number of events
over that sub-domain is shown  in Fig. \ref{f7} We see a strong rise towards
the lower   border of the kinematically allowed region. 
In other words the number of events increase with decreasing deuteron
energy. 
For the example of $T_{\pi}$ = 32.0
 MeV we display in Fig.\ref{f8} the dependence
of ${{ d \Gamma } \over {dT_{\pi} d T_{d}}}$ as a function of $T_{d}$. 

\begin{figure}
\centering
\input{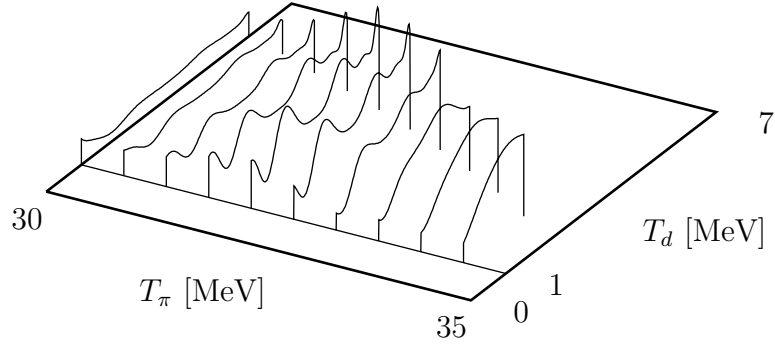}
\bigskip
\caption{The distribution ${{ d \Gamma ^{p+d}} \over {d T_{\pi} d T_{d}}}$ 
in a subdomain of Fig.\ref{f8} for $T_{d} \ge$ 1MeV. 
That interference pattern results from FSI and the scale of the figure is  a 
 factor 100 smaller than in  Fig. \ref{f7}. }
\label{f10}
\end{figure}

Finally we display in Fig. \ref{f10} ${{d \Gamma } \over {d T_{\pi } d T_{d
      }}} $ in the 
  same sub-domain as in Fig. \ref{f7} 
 but excluding deuteron energies $T_{d} \le$
  1MeV. The rate is down by about a factor 100 but it 
is more interesting, since it
  is generated just by final state interactions. We see  a rich interference 
pattern.

\section{Summary}
\label{secIV}

The mesonic decay of the hypertriton has been calculated using a hypertriton
wavefunction and 3N bound and scattering states, which are rigorous solutions
of the Faddeev equations.
Our results are based on the Nijmegen 93 NN potential and Nijmegen
hyperon-nucleon forces, which include the $\Lambda$-$\Sigma$ conversion.
The standard simple particle operator for free $\Lambda \to \pi + N$ decay
has been used. The interaction between the emitted $\pi^{-}$ and the nucleons
is neglected as well as Coulomb forces. We evaluated the partial decay rates
into the $^{3}$He + $\pi^{-}$, p + d + $\pi^{-}$ and p + p + n + $\pi^{-}$
channels. 
The corresponding rates for $\pi^{0}$ emission is given 
by isospin symmetry (if
valid). 
The total mesonic rate is 0.35 $\times$ 10$^{10}$ sec$^{-1}$, which is 92\% of the free
$\Lambda$-decay rate. 
If one adds the nonmesonic decay channels that number is 93\%. 
Thus the $^{3}_{\Lambda}$H lives a bit longer than a free
$\Lambda$.

In a previous work\cite{nonmesonic} we already discussed the nonmesonic decay
channels. Thus fairly complete theoretical predictions for all decay channels
of the hypertriton are available, which are based on modern forces and
rigorous 3-body calculations.

We also studied the momentum distribution of the emitted $\pi^{-}$. 
This  distribution appears to be nicely related to nucleon-deuteron scattering
initiated in the hypertriton (after decay of the $\Lambda$) as one can infer
from the relation between the p + d and p +  p + n channels.

The p + d + $\pi^{-}$ decay is visualised in form of a Dalitz plot. 
As expected the spectrum peaks for low energetic deuterons which are 
present in the weakly bound hypertriton.

Finally we compared the polarization of the outgoing proton in free
unpolarised $\Lambda$-decay to the polarisation of $^{3}$He in the unpolarised
hypertriton decay.
They turn out to be opposite in sign, which is explained by regarding a 
related process. This is the difference in $\pi^{-}$ rates for emission parallel
or antiparallel to the polarised $\Lambda$ and $^{3}_{\Lambda}$H
respectively. 
The two observables, the polarisation of the outgoing proton and the 
difference in  $\pi^-$ rates are equal up to a factor 2. The changes in sign 
results since a polarised $^3_\Lambda$H contains a polarised $\Lambda$ 
with  the spin direction pointing in the opposite direction of the 
polarisation of the hypertriton. 

Measurements of hypertriton decay properties would certainly be very useful to
test these predictions based on modern dynamics.

\acknowledgements

This work was supported by
the Research Contract \# 41324878 (COSY-044) of the Forschungszentrum J\"ulich,
the Deutsche Forschungsgemeinschaft and 
the Science and Technology Cooperation Germany-Poland under Grant No.~XO81.91.
The numerical calculations have been performed on the Cray T90 of the
H\"ochstleistungsrechenzentrum in J\"ulich, Germany.

\section{appendix}

\subsection{ Free $\Lambda $ decay } 

As is well known from text books, for instance 
\cite{Nishijima}, the total decay rate for $\Lambda \to p + \pi^{-} $ and
based on our notation (\ref{eo0})  is  
\begin{eqnarray}
\Gamma^{\Lambda \to p + \pi^{-} }_{rel} = { { a b ( f_{s} ^{2} a^{2} + f_{p}^{2} 
b^{2} ) } \over { 8 \pi M_{\Lambda} ^{3} } } 
\label{A1}
\end{eqnarray}
with 
\begin{eqnarray}
a &=& \sqrt{ (M_{\Lambda} + M_{p})^{2} - m_{\pi}^{2}} \cr
b &=& \sqrt{ (M_{\Lambda} - M_{p})^{2} - m_{\pi}^{2}} 
\end{eqnarray}
and 
\begin{eqnarray}
f_{s} &=& G_{F} m_{\pi} ^{2} A_{\pi} F(\vec k_{\pi}) \cr
f_{p} &=& G_{F} m_{\pi} ^{2} B_{\pi} F(\vec k_{\pi})
\end{eqnarray}
We choose the form factor $F(\vec k_{\pi}) $ as 
\begin{eqnarray}
F(\vec k_{\pi}) = { {  \Lambda_{\pi}^{2} - m_{\pi } ^{2} } \over
{ \Lambda _{\pi}^{2} + {\vec k}_{\pi } ^{2} } }
\end{eqnarray}
with the cut-off parameter $\Lambda_{\pi}$ = 1300 MeV used also in our 
nonmesonic decay calculation\cite{nonmesonic}.
The momentum $k_{\pi } $ is kinematically fixed as 
\begin{eqnarray}
k_{\pi} = \vert \vec k_{\pi } \vert = 
{ \sqrt{ (M_{\Lambda} ^{2} + M_{p} ^{2} - m_{\pi}^{2})^{2} - 4 M_{\Lambda}^{2}
M_{p}^{2} } \over { 2 M_{\Lambda}}}{\rm .}
\end{eqnarray}

The nonrelativistic calculation based on Eq. (\ref{eo}) yields 
\begin{eqnarray}
\Gamma^{\Lambda \to p + \pi^{-}}_{nonrel} = 
{ k_{\pi} \over { \pi}  } 
{M_{p} \over { M_{p} + \omega_{\pi} } } 
\left(  f_{s}^{2} +
    f_{p} ^{2}{ { k_{\pi} ^{2} } \over  {4 \overline{M}^{2}}} \right)
\label{A2}
\end{eqnarray}
Both results, Eqs. (\ref{A1}) and (\ref{A2}) agree within 1.7\%. 
Eqs.(\ref{A1})  and (\ref{A2}) lead to 
 life times of the $\Lambda$ particle 
 ${2 \over 3} [\Gamma^{\Lambda \to p + \pi^{-}}_{rel}]^{-1} $ 
= 2.579 $\times$ 10$^{-10} $ sec and ${2 \over 3} [\Gamma^{\Lambda \to p +
  \pi^{-}}_{nonrel}]^{-1} $ =  2.624 $\times$ 10$^{-10} $ sec, respectively,
 which compare well to the datum  
2.632 $\pm$ 0.020 $\times$ 10$^{-10}$ sec\cite{Particle}.


\subsection{Partial wave representation of the operator $O$}

We show the partial wave representation 
for the operator $O$ of Eq. (\ref{eo}) applied onto the hypertriton state.
One has 
\begin{eqnarray}
\langle p q \alpha m \vert O \vert \Psi _{^{3}_{\Lambda} H }
\, m_{^{3}_{\Lambda} H}  \rangle 
= \sum 
_{\alpha '} \int dp' {p '}^{2} \int dq ' {q'}^{2} 
  \langle p q \alpha m \vert O \vert p' q ' \alpha ' \rangle 
\langle p' q' \alpha ' \vert \Psi _{^{3}_{\Lambda} H} \, m_{^{3}_{\Lambda}H}
 \rangle
\end{eqnarray}
where m and $m_{^{3}_{\Lambda}H }$ are the  z-components of the 
outgoing 3N state and 
 the hypertriton, respectively. 
The expression  can be  separated into two terms 
\begin{eqnarray}
\langle p q \alpha \vert  O \vert \Psi _{^{3}_{\Lambda} H } \,
m_{^{3}_{\Lambda} H}  \rangle = 
i \sqrt{2}G_{F}m_{\pi}^{2} F(\vec k_{\pi})
A_{\pi} O_{A} + i\sqrt{2} G_{F}m_{\pi}^{2} F(\vec k_{\pi})
{ {B_{\pi}} \over {2 \overline M } }  O_{B} 
\end{eqnarray}
with 
\begin{eqnarray}
O_{A} &=& {1\over 2}\delta_{m m_{^{3}_{\Lambda}H}}
\sum_{{\alpha}'}
\sqrt{\hat J \hat I \hat I ' 
\hat \lambda '} \sqrt{\hat \lambda ' ! }
  (-)^{j} \delta _{l l'}
\delta_{s s'} \delta_{j j'} 
 \sum_{\lambda_{1}+\lambda_{2}=\lambda '} {q}^{\lambda_{1}}
( {2 \over 3} k_{\pi}) ^{\lambda_{2}} { 1 \over
  {\sqrt{(2\lambda_{1})!(2\lambda_{2})! }}} 
\cr
&{}&
\sum _{k} \hat k (-)^{k} g_{k \, \alpha '}(p,q,k_{\pi}) 
C(\lambda_{1} k \lambda , 0 0 )
\sum_{g} \sqrt{ \hat g} 
C( \lambda_{2} k g , 0 0)
\cr
&{}&
\left\{ \matrix{ g & \lambda & \lambda ' \cr
{ 1 \over 2 } & I ' & I \cr  } \right\}
\left\{ \matrix{ g & I & I' \cr 
j & { 1\over 2 } & J \cr} \right\} C(J\, g \, {1 \over 2} , m \, 0
\, m_{^{3}_{\Lambda}H} ) 
\label{A3}
\end{eqnarray} 
and 
\begin{eqnarray}
O_{B} &=& {\sqrt{6}\over 2} k_{\pi} \delta_{m m_{^{3}_{\Lambda}H}}
\sum_{\alpha '}
\delta _{l l'}
\delta_{s s'} \delta_{j j'}
(-)^{j +I' + {1 \over 2}}
\sqrt{\hat  J \hat I \hat I'  \hat \lambda \hat \lambda '} 
\sqrt{\hat \lambda ' ! }
\sum_{L} \hat L (-)^{L}
\left\{ \matrix{\lambda ' & L & 1 \cr
{1 \over 2} & {1 \over 2} & I ' \cr }\right\}
\cr
 &{}&
 \sum_{\lambda_{1}+\lambda_{2}=\lambda '} {q}^{\lambda_{1}}
( {2 \over 3} k_{\pi}) ^{\lambda_{2}} 
{ 1 \over
  {\sqrt{(2\lambda_{1})!(2\lambda_{2})! }}} 
\sum _{k} \sqrt{\hat k}  g_{k \, \alpha '}(p,q,k_{\pi}) 
C(\lambda_{1 } \lambda k, 0 0 )
\cr
&{}&
\sum_{h} \sqrt{\hat h} 
C( k  1 h , 0 0 )
\sum_{g} \sqrt{ \hat g } (-)^{g} 
C(\lambda_{2} h g , 0 0  )
\left\{ \matrix{  \lambda & g & L \cr
I' & {1 \over 2} & I \cr } \right\}
\cr 
&{}&
\left\{ \matrix{ \lambda_{1} & \lambda_{2} & \lambda ' \cr
k & h & 1 \cr
\lambda & g & L \cr } \right\}
\left\{ \matrix{ J & I & j \cr 
I' & { 1 \over 2} & g \cr } \right\} 
 C(J\, g \, {1 \over 2} , m \, 0
\, m_{^{3}_{\Lambda}H} )
\label{A4}
\end{eqnarray}
where
\begin{eqnarray}
g_{k \, \alpha ' }(p,q,k_{\pi}) = \int _{-1}^{1} dx P_{k} (x) 
{{ \langle p, \vert \vec q + {2 \over 3 } \vec k_{\pi} \vert \alpha ' \, 
\vert \,
\Psi _{^{3}_{\Lambda}H} \, 
m_{^{3}_{\Lambda}H} \rangle } \over { \vert \vec q + {2 \over 3 } \vec k_{\pi}
\vert  ^{\lambda '}} }
\end{eqnarray}
and 
where $P_{k}$ is a Legendre function depending on 
$x=\hat q \cdot \hat k_{\pi}$.
We use the notation $\hat z = 2 z + 1$.  
Also we 
assume the quantum axis to be  parallel to the $\vec k _{\pi}$ direction.

The isospin part of the matrix element is not included in (\ref{A3}) and
(\ref{A4}). It yields just the factor $\sqrt{2}$ for the $\pi^{-}$ transition.
Furthermore it leads to the requirement that the isospin of the two spectator
nucleons has to be zero and that only total isospin T=1/2 contributes.

\begin{table}
\caption{Partial and total mesonic and nonmesonic decay rates and
  corresponding life times.}
\begin{center}
\begin{tabular} {cccc}
channel & $\Gamma$ [sec$^{-1}$] & $\Gamma$/ $\Gamma_\Lambda$  &  
$\tau $=$\Gamma^{-1} $ [sec] 
\\ [4pt] \hline
 $^3$He +$ \pi ^-$ and $^3$H + $\pi^0$ & 0.14  $\times$ 10$^{10}$ & 0.37 & 0.71 $\times$ 10$^{-9}$ 
\\ [4pt] \hline
 d +p +$ \pi ^-$ and  d + n + $\pi^0$ &  0.21 $\times$ 10$^{10}$ & 0.55  & 0.47 $\times$ 10$^{-9}$
\\ [4pt] \hline
 p + p + n + $\pi^-$ and p + n + n +$\pi^0$ & 0.84 $\times$ 10$^7$ & 0.022 & 0.12 $\times$ 10$^{-6}$
\\ [4pt] \hline
all mesonic channels & 0.35  $\times$ 10$^{10}$ & 0.92 & 0.29 $\times$ 10$^{-9}$
\\ [4pt] \hline
\\ [4pt] \hline
  d + n  &     0.67  $\times$ 10$^7$ & 0.0018 &  0.15 $\times$ 10$^{-6}$
\\ [4pt] \hline
 p + n + n    &  0.57  $\times$ 10$^8$ & 0.015   & 0.18 $\times$ 10$^{-7}$
\\ [4pt] \hline
all nonmesonic channels & 0.64 $\times$ 10$^8$ & 0.017  & 0.16 $\times$ 10$^{-7}$
\\ [4pt] \hline
\\ [4pt] \hline
all channels          & 0.36  $\times$ 10$^{10}$  & 0.93  & 2.78 $\times$10$^{-10}$
\\ [4pt] \hline
exp.\cite{3}                 &                      &       & 2.64 +0.92 -0.54 
$\times $10$^{-10}$    
\\ [4pt] 
exp. (averaged) \cite{Congleton}                 &                  &       & 2.44 + 0.26 -0.22
$\times $10$^{-10}$    
    \\[4pt]
\end{tabular}
\end{center}
\label{t1}
\end{table}

\vfil \eject

\end{document}